\documentclass[twocolumn,aps,showpacs,preprintnumbers,amsmath,amssymb,unsortedaddress,superscriptaddress]{revtex4}
\usepackage{graphicx,epsf,epsfig,ulem}
\usepackage{bm}
\usepackage{subfigure}
\usepackage{color}
\usepackage{ulem}

\begin{document}

\title{Electronic transport driven spin-dynamics}
\bigskip
\author{L. Siddiqui}
\affiliation{School of Electrical and Computer Engineering, Purdue University, West Lafayette, Indiana 47907, USA}
\author{D. Saha}
\affiliation{Department of Electrical Engineering and Computer Science, University of Michigan, Ann Arbor, Michigan 48109, USA}
\author{S. Datta}
\affiliation{School of Electrical and Computer Engineering, Purdue University, West Lafayette, Indiana 47907, USA}
\author{P. Bhattacharya}
\affiliation{Department of Electrical Engineering and Computer Science, University of Michigan, Ann Arbor, Michigan 48109, USA}

\medskip
\widetext
\begin{abstract}
We propose a model to explore the dynamics of spin-systems coupled by exchange interaction to the conduction band electrons of a semiconductor material that forms the channel in a ferromagnet/semiconductor/ferromagnet spin-valve structure. We show that recent observation of the novel transient transport signature in a MnAs/GaAs/MnAs spin-valve structure with paramagnetic Mn impurities [D. Saha {\it{et al.}}, Phys. Rev. Lett., {\bf{100}}, 196603 (2008)] can be quantitatively understood in terms of current driven dynamical polarization of Mn spins. Using our model of spin polarized transport through Schottky barriers at the two ferromagnet/semiconductor junctions in a spin-valve structure and a dynamical equation describing the paramagnetic impurities coupled to conduction band electrons we explain the scaling behaviour of observed transient features such as the magnitude and time-scale with temperature.
\end{abstract}
\maketitle

\section{Introduction}
Spin-dynamics are finding wide interest in recent years as they usher in the prospect of finding their applications in quantum computing~\cite{kane,burkard}, data storage~\cite{gurney}, biological and medical applications~\cite{dunin,hoffmann}, etc. The success of these goals are contingent on being able to control and detect the spin degrees of freedom concerned. The succesful experimental demonstration of spin-injection from ferrmoagnetic contacts into semiconductors~\cite{ramsteiner,saha,jonker} and of controlling the spin-injection and extraction by spin-valve structures in semiconductors~\cite{saha} enables the possible control of surrounding spin-degrees of freedom by spin-polarized carriers inside the semiconductor strutures. Understanding the ensuing spin dynamics by the spin-polarized carrier is of immense importance to all novel electrically driven spin-based device concepts ranging from devices employing non-interacting spins~\cite{datta100} to those employing strongly interacting spins in magnets~\cite{salahuddin}.

\begin{figure}[]
\begin{center}
\includegraphics[width=0.45\textwidth]{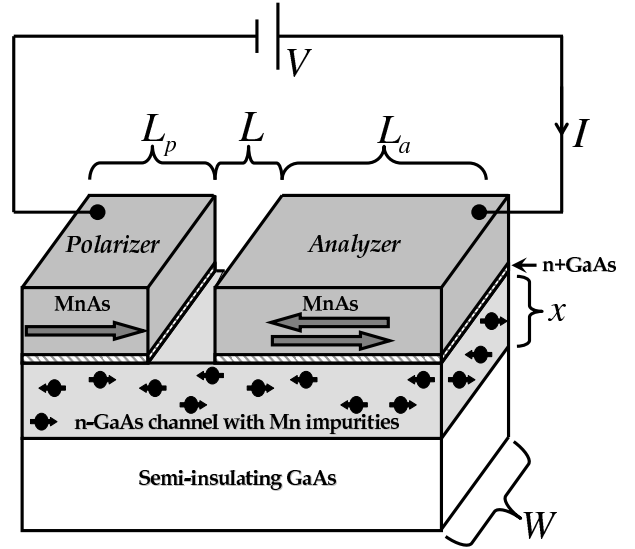}
\end{center}
\caption{Schematic structure of the Spin-charge Transducer (SCT). $L_{p(a)}$, $L$, $W$, and, $x$ denotes the length of the polarizer (analyzer), separation bewtween contacts, width of the entire structure, and, thickness of GaAs channel respectively. The magnetization of the analyzer can be reversed back and forth to to make it parallel (P) and anti-parallel (AP) to that of the polarizer.}
\label{sct_struc}
\end{figure}

We have developed a model to treat the non-equilibrium non-interacting spin dynamics driven by spin-polarized electrons injected into and extracted from a semiconductor channel via Schottky barriers between the semiconductor and ferromagnetic contacts arranged in a spin-valve configuration. By modeling the transport through these two Schottky barriers and solving them together in conjunction with a minimal transport model for the semiconductor, we can estimate the spin-flip current inside the semiconductor channel and thereby calculate the ensuing dynamical evolution of non-interacting spin-systems having exchange-interaction with the channel electrons.

\begin{figure}[]
\begin{center}
\includegraphics[width=0.45\textwidth]{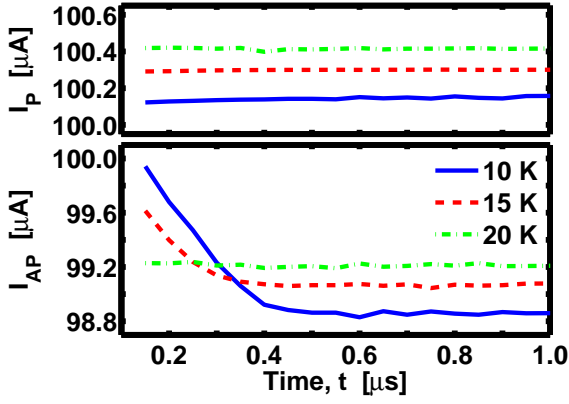}
\end{center}
\caption{Transient variation of terminal current $I$ for an applied bias $V=25$mV during parallel ($I_P$:{\it{top}}) and anti-parallel ($I_{AP}$:{\it{bottom}}) configuration at different temperature for an SCT structure with dimensions: $L_p=5\mu$m, $L=1\mu$m, $L_a=30\mu$m, $W=50\mu$m, and, $x=0.5\mu$m.}
\label{I_p_ap}
\end{figure}

We apply our model to analyze recent experimental observations~\cite{saha1} in MnAs/GaAs/MnAs lateral spin-valve structure with paramagnetic Mn impurities embedded in GaAs (fig.~\ref{sct_struc}). We attribute the origin of novel transient features such as the decay in the anti-parallel current over time (fig.~\ref{I_p_ap}) to dynamical polarization of paramagnetic Mn impurities ({\it{inset}} of fig.~\ref{Isf_Sz_t}) and can successfully explain the scaling of the transient characteristics such as the magnitude and time-constant with temperature (fig.~\ref{Isf_Sz_t}). In this regard we point to the suppressive role played by the other electronic spin-flip mechanisms in GaAs, which do not involve Mn impurities (fig.~\ref{Isf_Sz_t}), to reduce the effect at elevated temperature.

Our paper thus serves a two-fold purpose: (i) it identifies novel transport signatures pertaining to the underlying physical process of dynamical polarization of paramagnetic impurities embedded in a material carrying spin-polarized electrons which are coupled by exchange interaction to the impurities, and (ii) it identifies the effect of temperature and other electronic spin-relaxation mechanisms on the mentioned signatures.

\section{Theory and Calculation}

\begin{figure}[]
\begin{center}
\includegraphics[width=0.45\textwidth]{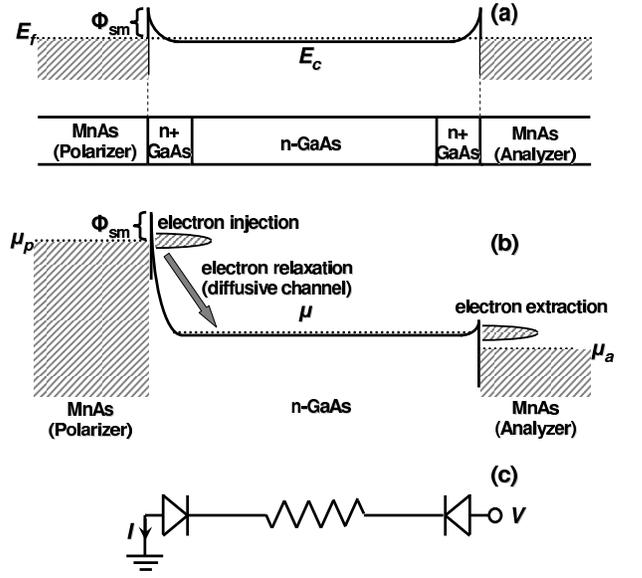}
\end{center}
\caption{{\it{(a)}} Equillibrium band diagram: the GaAs channel is coupled to the MnAs contacts through Schottky barriers. $E_f$ and $E_c$ are the equillibrium chemical potential and conduction band edge of the GaAs respectively. {\it{(b)}} Band diagram under bias: The polarizer Schottky barrier is forward biased and the Analyzer Schottky barrier is reverse biased. The electrons are being injected into (extracted from) the GaAs channel, maintaining a chemical potential $\mu_{chan}$, from (to) the polarizer (analyzer) MnAs conatact having chemical potential $\mu_p$ ($\mu_a$) {\it{(c)}} Equivalent circuit diagram for SCT with non-magnetized contacts: the applied voltage $V$ drives a terminal current $I$ through two diodes connected in series by a resistor.}
\label{band_ckt}
\end{figure}

\subsection{Spin transport through the SCT}
\noindent The basic electronic transport mechanism through the SCT is that electrons tunnel through two back-to-back Schottky barriers at the two MnAs/GaAs junctions, which are connected via a diffusive semiconductor channel (fig.~\ref{band_ckt}). To model the tunneling Schottky barrier the magnetized MnAs contacts the model equations in refs.~\cite{sze,stratton} needs to be modifeid to consider that (i) the magnetized MnAs density of states and Fermi velocity for the up and the down electrons are different around the chemical potential, and (ii) the up-spin and the down-spin electrons might see different electrochemical potentials $V_{up(a)}$ and $V_{dp(a)}$ respectively while tunneling across the forward (reversed) biased polarizer (analyzer) junction but the same electrostatic potential $V_{esp(a)}$ as the electrostatic potential depends on the total charge density. We introduce a contact polarization parameter $P_c$ to consider the first effect which implies that even if both the up-spin and down-spin electrons see the same electrochemical potential while tunneling across the barrier, the ratio of majority to minority spin-component of the current is, $(1+P_c):(1-P_c)$. In all our calculations we will follow the convention that when the two magnetic MnAs contacts have parallel magnetization the majority (will be denoted by $M$) and minority (will be denoted by $m$) spin-components of both the contacts are up-spin (will be denoted by $u$) and down-spin respectively (denoted by $d$). Then it follows that when the contacts have anti-parallel magnetization the majority and the minority spin-components of the analyzer will be the opposite of that of the parallel case (i.e. down-spin and up-spin respectively) as it's magnetization is reversed during the experiment. The second effect will directly modify the tunnel current expressions in refs.~\cite{sze,stratton} in such a way that the electrostatic potential essentially enters the terms that comes from the shape of the barrier and the two electrochemical potentials enters the terms that involves the position of the Fermi levels. After making these modifications we finally get the expressions for majority and minority spin-component of current through the polarizer (analyzer) $I_{p(a)M}$ and $I_{p(a)m}$ respectively:
\begin{eqnarray}
I_{p(a)M}=0.5(1+P_{c})L_{p(a)}WJ^{RB(FB)}_{sch}(V_{esp(a)},V_{Mp(a)})\label{eq91}\\
I_{p(a)m}=0.5(1-P_{c})L_{p(a)}WJ^{RB(FB)}_{sch}(V_{esp(a)},V_{mp(a)})\label{eq92}
\label{sch_curr_maj_min}
\end{eqnarray} where $W$ is the width of the n-GaAs channel (fig.~\ref{sct_struc}) and $L_{p(a)}$ is the length of the polarizer (analyzer) and $(M,m)=(u,d)$ for both the contacts during parallel magnetization and $(M,m)=(d,u)$ only for the analyzer during anti-parallel magnetization. The current density is given by:
\begin{eqnarray}
&&J^{FB(RB)}_{sch}(V_{esa(p)},V_{ya(p)})=\frac{A^*e^{-b^{FB(RB)}}_1}{(c^{FB(RB)}_1k_BT)^2}\nonumber\\
&&\times\frac{\pi c^{FB(RB)}_1k_BT}{sin(\pi c^{FB(RB)}_1k_BT)}\{1-e^{-c^{FB(RB)}_1q|V_{ya(p)}|}\}\\\label{sch_curr_mag}
&&A^*=4\pi m^*q(k_BT)^2/(2\pi\hbar)^3\nonumber
\end{eqnarray} where, $y\in\{M,m\}$, $k_B$ is the Boltzman constant, $q$ is the electronic charge, and $T$ is the temperature. Now for forward bias:
\begin{eqnarray}
b^{FB}_1&=&(\phi_{sm}-qV_{esa})/E_{00}\label{eq101}\\
c^{FB}_1&=&\frac{1}{2E_{00}}log\frac{4(\phi_{sm}-qV_{esa})}{E_{fc1}+q(V_{xa}-V_{esa})}\label{eq102}
\end{eqnarray} and for reverse bias:
\begin{eqnarray}
b^{RB}_1&=&\frac{1}{E_{00}}\left[\phi^{1/2}_{sm}(\phi_{sm}+qV_{esp})^{1/2}\right.\nonumber\\
&&\left.-qV_{esp}log\frac{(\phi_{sm}+qV_{esp})^{1/2}+\phi^{1/2}_{sm}}{(qV_{esp})^{1/2}}\right]\label{eq103}\\
c^{RB}_1&=&\frac{1}{E_{00}}log\frac{(\phi_{sm}+qV_{esp})^{1/2}+\phi^{1/2}_{sm}}{(qV_{esp})^{1/2}}\label{eq104}
\end{eqnarray} Here, $E_{00}=2q\sqrt{N_{D1}/2\epsilon}/\alpha$ with $\alpha=2\sqrt{2m^*}/\hbar$. $\hbar$ is the Plank's constant, $m^*$ is the effective mass of GaAs conduction band electrons, $\epsilon$ is the dielectric permittivity of GaAs, $N_{D1}$ is the doping density of the n+GaAs region (figs.~\ref{sct_struc} and~\ref{band_ckt}(a)), $E_{fc1}$ is the position of the equillibrium Fermi energy of the n+GaAs above it's conduction band-edge, and $\phi_{sm}$ is the Schottky barrier height at equillibrium, which will be used as a fitting parameter. In the expressions above all the electrochemical and electrostatic potential are not completely independent due to the restriction of charge neutrality. Assuming that the difference between the up-spin and down-spin charge density to be smaller enough than the total charge density we can write:
\begin{eqnarray}
V_{esa(p)}=(V_{Ma(p)}+V_{ma(p)})/2
\label{estat_echem}
\end{eqnarray} Finally, electrostatic potentials across the polarizer, analyzer and the channel ($V_{chan}$) should add up to applied voltage $V$ and the currents through the polarizer ($I_p$), analyzer ($I_a$) and channel ($I_{chan}$) should all be equal to the current in the external circuit:
\begin{eqnarray}
V=V_{esp}+V_{esa}+V_{chan}\label{kvl_mag}\\
I=I_{p}=I_{a}=I_{chan}\label{kcl_mag}\\
I_{p}=I_{pM}+I_{pm}\label{pol_curr_mag}\\
I_{a}=I_{aM}+I_{am}\label{an_curr_mag}
\end{eqnarray} The relation between $V_{chan}$ and $I_{chan}$ is given by:
\begin{eqnarray}
I_{chan}&=&\frac{Wx}{L+L_a/2+L_p/2}qN_{D2}\nu V_{chan}\label{curr_chan_no_mag}
\end{eqnarray} where, $x$, $N_{D2}$, and $\nu$ are the thickness, doping density and mobility of the nGaAs channel respectively (figs.~\ref{sct_struc}) and $L$ is the separation between the contacts.

From equations \ref{sch_curr_mag},~\ref{eq103} and~\ref{eq104} we see that the electrochemical potential across the reverse biased junction only enters through $\{1-e^{-c^{FB(RB)}_1q|V_{ya(p)}|}\}$ which becomes $1$ for the reverse biased polarizer voltage of interest. So we can assume that for the reverse biased junction (i.e. polarizer) the exact value of the electrochemical potential $V_{yp}$ is of no consequence and will always be set equal to electrostatic potential $V_{esp}$ in all the calculations to follow, i.e.
\begin{eqnarray}
V_{Mp}=V_{mp}=V_{esp}
\label{rb_approx}
\end{eqnarray}

\subsection{Estimating the spin-flip current}
\begin{figure}[]
\begin{center}
\includegraphics[width=0.45\textwidth]{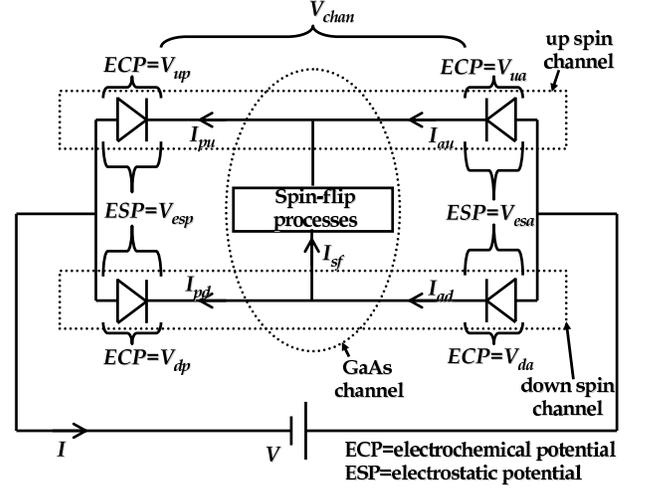}
\end{center}
\caption{Equivalent circuit diagram for the SCT with magnetized contacts}
\label{sct_eq_ckt}
\end{figure}

\noindent As the majority and minority electron spins enter and flow through the channel material, different spin-flip mechanisms can transfer some electrons from the up-spin stream to the down-spin stream and vice-verca. As a result, the extracted up(down)-spin current $I_{au(d)}$ might not be the same as the injected up(down)-spin current $I_{pu(d)}$. The difference is what we will call the spin-flip current:
\begin{eqnarray}
I_{sf}=I_{pu}-I_{au}=I_{ad}-I_{pd}
\label{spin_flip_curr}
\end{eqnarray} A schematic circuit diagram with different current component discussed so far is shown in Fig.~\ref{sct_eq_ckt}.

\noindent {\it{Parallel configuration:}} We will start by analyzing the data for the parallel configuration. In this configuration we can rewrite the eqs.~\ref{eq91} and~\ref{eq92}:
\begin{eqnarray}
I_{pu}=0.5(1+P_c)L_pWJ^{RB}_{sch}(V_{esp},V_{up}=V_{esp})\label{eq1}\\
I_{pd}=0.5(1-P_c)L_pWJ^{RB}_{sch}(V_{esp},V_{dp}=V_{esp})\label{eq2}\\
I_{au}=0.5(1+P_c)L_aWJ^{FB}_{sch}(V_{esa},V_{ua})\label{eq3}\\
I_{ad}=0.5(1-P_c)L_aWJ^{FB}_{sch}(V_{esa},V_{da})\label{eq4}
\end{eqnarray} In all following we will not consider the negligible time-dependence of parallel current data and will assume that the it has a constant value over time. Then it follows that there is no significant spin-flip dynamics taking place during parallel configuration and we can assume spin-flip current to be zero. For zero spin-flip current we get from eq.~\ref{spin_flip_curr},~\ref{eq1} and~\ref{eq2}:
\begin{eqnarray}
\frac{I_{pu}}{I_{pd}}=\frac{I_{au}}{I_{ad}}=\frac{1+P_c}{1-P_c}\label{eq5}
\end{eqnarray} Now from eqs.~\ref{estat_echem},~\ref{eq3},~\ref{eq4} and~\ref{eq5} we get:
\begin{eqnarray}
V_{ua}=V_{da}=V_{esa}\label{eq6}
\end{eqnarray} Substituting~\ref{eq6} and~\ref{eq1} -~\ref{eq4} into~\ref{kcl_mag} we find that the parallel current:
\begin{eqnarray}
I_P=L_pWJ^{RB}_{sch}(V_{esp},V_{esp})=L_aWJ^{FB}_{sch}(V_{esa},V_{esa})\label{eq131}
\end{eqnarray} is independent of$P_c$ and only parametrically depends on $\phi_{sm}$. For a particular value of parallel current $I$ and a particular value of $\phi_{sm}$ we can solve eq.~\ref{eq131} to determine $V_{esp}$ and $V_{esa}$. From the knowledge of $I_{chan}=I$ we can determine $V_{chan}$ from eq.~\ref{curr_chan_no_mag}. Subsequently, we can calculate $V$ from eq.~\ref{kvl_mag} for certain values of $I$ and $\phi_{sm}$. Conversely, by aplying iterative solution technique on these steps we can determine $\phi_{sm}$ from the experimental value of $I$ and $V$ for parallel case and the calculated value is $\phi_{sm}\approx 0.6$ which is very close to the known value in literature~\cite{garcia} and fits the data for all the experimental temperatures.
\begin{figure}[]
\begin{center}
\includegraphics[width=0.45\textwidth]{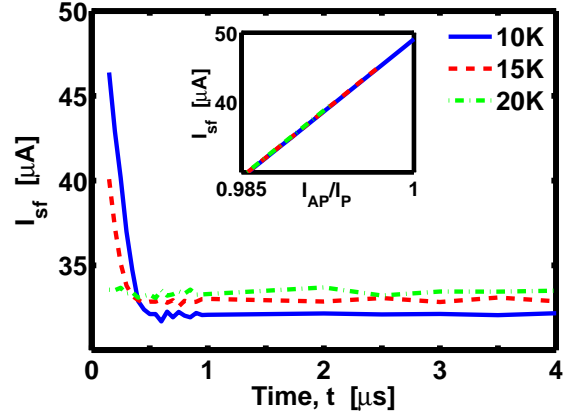}
\end{center}
\caption{Time variation of spin-flip current, $I_{sf}$ at different temperaures during antiparallel configuration. {\it{inset}}: Variation of spin-flip current $I_{sf}$ with antiparallel terminal current normalized by parallel terminal current ($I_{AP}/I_{P}$) at those temperatures. All the results are for same the voltage and dimension values as that of fig.~\ref{I_p_ap} and $P_c=0.49$~\cite{mazin}. The values of the other parameters used for this calculation are: $N_{D1}=1\times10^{19}$cm$^{-3}$, $E_{fc1}=400$meV, $\phi_{sm}\approx 0.6$eV~\cite{garcia}, and, $N_{D2}=8.6\times10^{16}$cm$^{-3}$.}
\label{Isf_t_Iap}
\end{figure}

\noindent {\it{Antiparallel configuration:}} The prominent transient behaviour of the AP current indicates significant spin-flip dynamics. With the value of $\phi_{sm}$ estimated from the P configuration data we will now proceed to estimate the spin-flip current variation over time. We first re-write eqs.~\ref{eq91} and~\ref{eq92} for the AP case:
\begin{eqnarray}
I_{pu}=0.5(1+P_c)L_pWJ^{RB}_{sch}(V_{esp},V_{up}=V_{esp})\label{eq201}\\
I_{pd}=0.5(1-P_c)L_pWJ^{RB}_{sch}(V_{esp},V_{dp}=V_{esp})\label{eq202}\\
I_{au}=0.5(1-P_c)L_aWJ^{FB}_{sch}(V_{esa},V_{ua})\label{eq203}\\
I_{ad}=0.5(1+P_c)L_aWJ^{FB}_{sch}(V_{esa},V_{da})\label{eq204}
\end{eqnarray} By adding eqs.~\ref{eq201} and~\ref{eq202} we get the AP terminal current:
\begin{eqnarray}
I_{AP}=I_{pu}+I_{pd}=L_pWJ^{RB}_{sch}(V_{esp},V_{esp})
\label{eq205}
\end{eqnarray} Now for a particular value of AP terminal current $I_{AP}$ we determine the value of electrostatic potential $V_{esp}$ using eq.~\ref{eq205}. Then for a particular value of contact polarization $P_c$ we determine the up-spin and down-spin components of the current ($I_{pu}$ and $I_{pd}$ respectively) injected through the polarizer terminal from:
\begin{eqnarray}
I_{pu}=0.5(1+P_c)I\label{eq206}\\
I_{pd}=0.5(1-P_c)I\label{eq207}
\end{eqnarray} We next determine the two spin-components $I_{au}$and $I_{ad}$ of analyzer current from~\ref{spin_flip_curr}. At this stage we know the values of $\phi_{sm}$ from parallel data, the currents $I_{au}$ and $I_{ad}$ for a particular value of $P_c$ and solve three eqs.~\ref{estat_echem},~\ref{eq203} and~\ref{eq204} for three unknown variables $V_{esa}$, $V_{ua}$ and $V_{pa}$. Finally we determine $V_{chan}$ from eq.~\ref{curr_chan_no_mag} and~\ref{kcl_mag}. With different potentials $V_{esp}$, $V_{esa}$ and $V_{chan}$ known we readily determine the applied voltage $V$ from eq.~\ref{kvl_mag}. By execution of all the steps mentioned above we finally deduce the value of $V$ for arbitrary values of $P_c$, $I_{sf}$, and $I$ using the value of $\phi_{sm}$ determined from the parallel data. Again, by iterative solution technique we can use these steps to determine the value of $I_{sf}$ for a particular value of $I$, $V$, and $P_c$. The results of such a calculation are shown in the inset of fig.~\ref{Isf_t_Iap}. Using these results (in the inset of fig.~\ref{Isf_t_Iap}) and the values of the experimental antiparallel current (fig.~\ref{I_p_ap}) we determine the time variation of $I_{sf}$ which is reported in fig.~\ref{Isf_t_Iap}.

\subsection{Impurity dynamics}
\noindent The Mn impurities in GaAs are compensated. Hence, they have a single negative charge and are in $S=5/2$ spin state \cite{almeleh}. We will estimate the impurity spin polarization in terms of the average z-component value $\langle S_z\rangle$ of impurity spins:
\begin{eqnarray}
\langle S_z\rangle&=&\frac{5}{2}F_{+5/2}+\frac{3}{2}F_{+3/2}+\frac{1}{2}F_{1/2}\nonumber\\
&-&\frac{1}{2}F_{-1/2}-\frac{3}{2}F_{-3/2}-\frac{5}{2}F_{-5/2}
\label{def_sz}
\end{eqnarray} where $F_n$ is the probability of an impurity atom being in $S_z=n$ state. When the impurity-spins are exchange coupled to the electron-spins the ensuing electron spin-flip current $I^{Mn}_{sf}$ can be written as (by extending the equations in \cite{datta101} to S=5/2 system):
\begin{eqnarray}
I^{Mn}_{sf}&=&\frac{q}{h}2\pi J^2N_I\int D_u(E)D_d(E)\left[\left(F_{-5/2}+F_{-3/2}+\right.\right.\nonumber\\
&&\left.\left.F_{-1/2}+F_{+1/2}+F_{+3/2}\right)f_u(E)\{1-f_d(E)\}-\right.\nonumber\\
&&\left.\left(F_{-3/2}+F_{-1/2}+F_{+1/2}+F_{+3/2}+F_{+5/2}\right)\times\right.\nonumber\\
&&\left.f_d(E)\{1-f_u(E)\}\right]dE
\label{sflip_5by2}
\end{eqnarray} where, $D_{u(d)}$ and $f_{u(d)}$ are the density of states (DOS) and average energy distribution functions of up(down)-spin electrons inside the channel, $N_{I}$ is the total no. of impurities, and, $h=2\pi\hbar$. As the impurities get spin-polarized their dynamics can be described as:
\begin{eqnarray}
\frac{d}{dt}\left[\begin{array}{c}
F_{+5/2}\\F_{+3/2}\\F_{+1/2}\\F_{-1/2}\\F_{-3/2}\\F_{-5/2}
\end{array}\right]=\frac{\gamma_I}{6}\left[\begin{array}{c}
1\\1\\1\\1\\1\\1
\end{array}\right]+\mathbf{\Gamma}\left[\begin{array}{c}
F_{+5/2}\\F_{+3/2}\\F_{+1/2}\\F_{-1/2}\\F_{-3/2}\\F_{-5/2}
\end{array}\right]
\label{imp_5by2_dyn}
\end{eqnarray} Here, $\gamma_I$ is the relaxation rate of the impurity-spins independent of electrons and
\begin{widetext}
\begin{eqnarray}
\mathbf{\Gamma}=\left[\begin{array}{cccccc}
-(B+\gamma_I) & A & 0 & 0 & 0 & 0\\
B & -(A+B+\gamma_I) & A & 0 & 0 & 0\\
0 & B & -(A+B+\gamma_I) & A & 0 & 0\\
0 & 0 & B & -(A+B+\gamma_I) & A & 0\\
0 & 0 & 0 & B & -(A+B+\gamma_I) & A\\
0 & 0 & 0 & 0 & B & -(A+\gamma_I)
\end{array}\right]
\label{imp_5by2_mat}
\end{eqnarray} 
\end{widetext} where,
\begin{eqnarray}
A=\frac{2\pi}{h}2\int J^2D_uD_df_u(1-f_d)dE\nonumber\\
B=\frac{2\pi}{h}2\int J^2D_uD_df_d(1-f_u)dE\nonumber
\label{coeff_AB}
\end{eqnarray} Now from eqs.~\ref{def_sz},~\ref{sflip_5by2}, and,~\ref{imp_5by2_dyn} we can write:
\begin{eqnarray}
\frac{1}{qN_I}I^{Mn}_{sf}(t)=\frac{d\langle S_z\rangle}{dt}+\gamma_I\langle S_z(t)\rangle
\label{sflip_dyn}
\end{eqnarray} As the spin-lattice relaxation of Mn impurity spins are negligible within the experimental temparature range \cite{almeleh} $\gamma_I\approx 0$. Consequently, it follows that $I^{Mn}_{sf}(t\to\infty)=0$ i.e. the spin-flip current component involing Mn impurities have zero contribution to the steady state spin-flip current. It is the other component of the spin-flip current $I^{so}_{sf}$, which involves conduction band electron spin-flip due to the spin-orbit interaction of GaAs, that  gives rise to steady state spin-flip current: $I^{so}_{sf}=I_{sf}(t\to\infty)$ and is always flowing as a `background' spin-flip current. Therefore we can find the spin-flip current due to Mn impurities from the results of fig.~\ref{Isf_t_Iap} as $I^{Mn}_{sf}(t)=I_{sf}(t)-I_{sf}(t\to\infty)$ and the results are shown in fig.~\ref{Isf_Sz_t}.

The GaAs channel being diffusive, we can assume that the energy distribution function of the up(down) electrons $f_{u(d)}$ to be Fermi functions with temperature $T$ and chemical potentials $\mu_{u(d)}$. We assume that $\mu_u-\mu_d$ is small and does not vary significantly over time. Then we have, due to charge neutrality condition inside the channel, the constraint $(\mu_u+\mu_d)/2=E_{fc2}$, where $E_{fc2}$ is the Fermi level of the n-GaAs channel corresponding to the doping of the channel (and not to be confused with $E_{fc1}$ which is the Fermi level of n+GaAs region). Also the GaAs having spin-degenerate bands (i.e. $D_u(E)=D_d(E)$) we can consider the density of states inside the channel to be uniform over the small energy window and define $D=D_u(E_{fc2})=D_d(E_{fc2})$. With these simplifications we can calculate the time variation of $I^{Mn}_{sf}$ and $\langle S_{z}\rangle$ from eqs.~\ref{def_sz},~\ref{sflip_5by2}, and,~\ref{imp_5by2_dyn} for a particular value of $JD$ and $\mu_u-\mu_d$ with the values of $E_{fc2}$ and $T$ set by the experimental conditions. Such a calculation was performed by varying $JD$ and $\mu_u-\mu_d$ to match $I^{Mn}_{sf}$ calculated from fig.~\ref{Isf_t_Iap} for $T=10$K and the results are shown in fig.~\ref{Isf_Sz_t}. As the temperature is raised the value of $JD$ should, in principle, remain unchanged and so a similar calculation was performed by varying only $\mu_u-\mu_d$ and keeping the value of $JD$ unchanged to the previously determined value to successfully match the $I^{Mn}_{sf}$ calculated from fig.~\ref{Isf_t_Iap} for $T=15$K. These results are also shown in fig.~\ref{Isf_Sz_t}.

\section{Discussion}
\begin{figure}[]
\begin{center}
\includegraphics[width=0.45\textwidth]{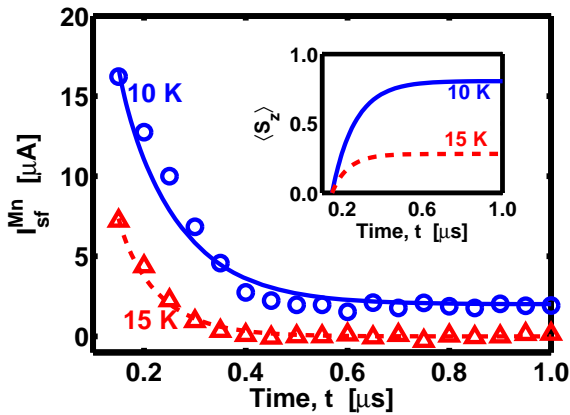}
\end{center}
\caption{Transient variation of $I^{Mn}_{sf}$ calculated from the impurity dynamics (solid and dashed lines) and from the calculation corresponding to fig.~\ref{Isf_t_Iap} (circles and triangles) (the results corresponsing to $10$K has been shifted upwards by $2\mu A$. {\it{Inset}}: Transient variation of the average z-cmponent $\langle S_z\rangle$ of Mn impurity spin at different temperatures. The parameters values for calculation corresponding to the impurity dynamics are: $JD=4.8\times10^{-3}$ and $\mu_u-\mu_d=0.25$mV ($0.125$mV) for $T=10$K ($15$K).}
\label{Isf_Sz_t}
\end{figure}

Comparing the results in fig.~\ref{Isf_Sz_t} at different temperatures we find that the difference between the chemical potentials of the up-spin and down-spin electrons $\mu_u-\mu_d$ decreases with increasing temperature. This is to be expected as the spin-orbit component of the spin-flip current $I^{so}_{sf}$ due to electron spin-relaxation mechanism in GaAs is known to increase with increasing temperature from magnetoresistance measurements on spin-valves~\cite{saha}. Such increase in $I^{so}_{sf}$ `loads' the chemical potential difference between up-spin and down-spin electrons $\mu_u-\mu_d$ that acts as a source to this component. As $\mu_u-\mu_d$ decreases with increasing temperature the initial value of $I^{Mn}_{sf}$, which is proportional to $\int(f_u-f_d)dE\thicksim(\mu_u-\mu_d)$ (eq.~\ref{sflip_5by2_1}), is expected to decrease according to our theoretical analysis and is also observed in the experiment. On the other hand, it can be shown that the time-scale of the transient variation, according eq.~\ref{imp_5by2_dyn}, is inversely proportional to $\int\{f_u(1-f_d)+f_d(1-f_u)\}dE$ which can be shown to be equal to $2k_BT$ for such small value of $\mu_u-\mu_d$. Hence according to our theory we expect the time-scale to be inversely proportional to the temperature, which is observed in the experiment: the value of time constant changed from $120$ns to $80$ns (their ratio being $3:2$) as the temperature was changed from $10$K to $15$K (their ratio being $2:3$).

We estimate the spin-relaxation time of the GaAs conduction band electrons, independent of Mn impurities, from the steady state spin-flip current values in fig.~\ref{Isf_t_Iap} ($I^{so}_{sf}$) and the chemical potential difference between up-spin and down-spin electrons ($\mu_u-\mu_d$ in fig.~\ref{Isf_Sz_t}) by determining the spin-flip conductance $g_{so}=I^{so}_{sf}/(\mu_u-\mu_d)$ and then equating the conductance value to the expression $\frac{q^2}{h}DWx(L+L_p+L_a)\frac{\hbar}{\tau_{so}}$. The estimated spin-relaxation time $\tau_{so}$ of the electrons independent of Mn impurities is found to be $2$ns ($1$ns) for T=$10$K ($15$K), which are very close to to the reported values~\cite{dzhioev}.

Furthermore, the strength of the exchange interaction ($J/a^3_{0}$) between conduction band electrons and the Mn impurities is estimated to be in the order of $1$eV (from the value of $JD$ in fig.~\ref{Isf_Sz_t} with the value of $D$ calculated using $E_{fc2}=10$meV corresponding to channel doping) which is an order of magnitude larger than the estimated s-d exchange interaction at band-edge in~\cite{dietl} and two orders of magnitude larger than the same in~\cite{awschalom}.

\section{Conclusion}

We have studied electronic transport driven non-equilibrium spin-dynamics arising from exchange interaction between the conduction band electrons and paramagnetic Mn spin impurities embedded in the GaAs channel of a MnAs/GaAs/MnAs spin-valve. We believe that we have sufficiently and uniquely elucidated the underlying physical processes leading to the experimental observations made in Ref.~\cite{saha1} as we can match several experimental features (parallel current values at 3 temperature settings, initial values and time-constants of $I^{Mn}_{sf}(t)$ at 2 temperature settings extracted from the experiment) using a few fitting parameters ($\phi_{sm}$, $JD$ and 2 values of $\mu_{u}-\mu_{d}$ at 2 different temperatures). From our calculations we argue that spin polarized carriers can significantly polarize the surrounding spin-systems and that the dynamical polarization has an impact on transport characteristics. We also observe that the other electronic spin-relaxation mechanisms suppresses the mentioned effect. In this regard, we would argue that using semiconductor materials having larger electronic spin-relaxation time such as Si, its spin-relaxation time being at least two orders of magnitude larger than GaAs~\cite{appelbaum}, can improve the polarization of the impurity spins by reducing the `loading' of $\mu_u-\mu_d$ given that the spin-injection efficiency does not suffer. Nevertheless, the degree of polarization acheived by such a small current is quite remarkable if we consider the fact that, to attain such polarization one would have to otherwise apply a large magnetic field (e.g. $\sim 2$ T to get a polarization of $\sim 0.8$ for an $S=5/2$ system at $10$K). This fact leads us to suggest that spin-polarized electronic current can emerge as a good candidate for all the experimental investigations and applications that necessitates manipulation of various spin-systems. An interesting extension of this work would be to study the effect of transport on weakly-interacting spin-systems such as a magnet near it's Curie temperature, which might have a prospect for a realization of electrically controlled magnetic ordering.

\appendix
\section{Derivation of eq.~\ref{sflip_dyn}}
Applying the sum rule:
\begin{eqnarray}
\sum_{n=-5/2}^{+5/2}F_{n}=1
\label{spin_sum}
\end{eqnarray} to eq.~\ref{sflip_5by2} we get the result:
\begin{eqnarray}
I^{Mn}_{sf}&=&\frac{q}{h}2\pi J^2N_I\int D_uD_d\left\{(f_u-f_d)+F_{-5/2}f_d(1-f_u)\right.\nonumber\\
&&\left.-F_{+5/2}f_u(1-f_d)\right\}dE
\label{sflip_5by2_1}
\end{eqnarray} Now from the definition in eq.~\ref{def_sz} we can write:
\begin{eqnarray}
\frac{d\langle S_z\rangle}{dt}=\left[\begin{array}{c}
+5/2\\+3/2\\+1/2\\-1/2\\-3/2\\-5/2
\end{array}\right]^T\frac{d}{dt}\left[\begin{array}{c}
F_{+5/2}\\F_{+3/2}\\F_{+1/2}\\F_{-1/2}\\F_{-3/2}\\F_{-5/2}
\end{array}\right]\nonumber
\end{eqnarray} to which we apply eqs.~\ref{imp_5by2_dyn} and \ref{imp_5by2_mat} and get:
\begin{eqnarray}
&&\frac{d\langle S_z\rangle}{dt}=(A-B)-AF_{+5/2}+BF_{-5/2}\nonumber\\
&=&\frac{1}{h}2\pi J^2\int D_uD_d\left\{(f_u-f_d)+F_{-5/2}f_d(1-f_u)\right.\nonumber\\
&&\left.-F_{+5/2}f_u(1-f_d)\right\}dE-\gamma_I\langle S_z\rangle\nonumber
\end{eqnarray} Substituting eq.~\ref{sflip_5by2_1} in the above equation we get the desired result of eq.~\ref{sflip_dyn}.

\end{document}